\documentclass[showpacs,preprintnumbers,showkeys,superscriptaddress,twocolumn]{revtex4}  
\usepackage{amssymb}
\usepackage{graphicx}
\usepackage{dcolumn}
\usepackage{bm}
\usepackage{appendix}
\usepackage{epstopdf}

\def\eqref#1{Eq.~(\ref{eq:#1})}

\begin{document}

\title{Variational Principle Directly on the Coherent Pair Condensate. I. the BCS Case}
\author{L. Y. Jia}  \email{liyuan.jia@usst.edu.cn}
\affiliation{Department of Physics, University of Shanghai for
Science and Technology, Shanghai 200093, P. R. China}

\date{\today}


\begin{abstract}

We propose a scheme to perform the variational principle directly on the coherent pair condensate (VDPC). The result is equivalent to that of the so-called variation after particle-number projection, but now the particle number is always conserved and the time-consuming projection is avoided. This work considers VDPC+BCS. We derive analytical expressions for the average energy and its gradient in terms of the coherent pair structure. In addition, we give analytically the pair structure at the energy minimum, and discuss its asymptotic behavior away from the Fermi surface, which looks quite simple and allows easy physical interpretations. The new algorithm iterates these pair-structure expressions to minimize energy. We demonstrate the new algorithm in a semi-realistic example using the realistic $V_{{\rm{low}}{\textrm{-}}k}$ interaction and large model spaces (up to $15$ harmonic-oscillator major shells). The energy can be minimized to practically arbitrary precision. The result shows that the realistic $V_{{\rm{low}}{\textrm{-}}k}$ interaction does not cause divergences in the pairing channel, although tiny occupation numbers (for example smaller than $10^{-5}$) contribute to the energy (by a few tens of keV). We also give analytical expressions for the gradient of energy with respect to changes of the canonical single-particle basis, which will be necessary for the next work in this series: VDPC+HFB.

\end{abstract}


\vspace{0.4in}

\maketitle


\section{Introduction}

Many nuclear structure theories start from a mean-field picture \cite{Bender_2003}. The choices for the mean field can be either phenomenological or microscopic. The phenomenological type includes the widely used spherical harmonic oscillator, Woods-Saxon, and Nilsson mean field. The microscopic theory is the Hartree-Fock (HF) method.

The pairing correlation has long been recognized \cite{Bohr_1958} and influences practically all nuclei across the nuclear chart \cite{Bohr_book, Ring_book}. To incorporate pairing into the mean field, we introduce quasiparticles and use for example Nilsson + BCS, HF + BCS, or Hartree-Fock-Bogoliubov (HFB) theory. These theories are examples of the variational principle, where the trial wavefunction is the quasiparticle vacuum. In the BCS case only the pair structure (occupation probability) is varied, whereas in the HFB case the pair structure is varied together with the canonical basis.

However, the BCS or HFB theory has the drawback of breaking the exact particle number \cite{Ring_book}. Only the average particle number is guaranteed by the chemical potential. Effectively, we replace the original micro-canonical ensemble by the grand-canonical ensemble (at zero temperature). The two ensembles are equivalent in the thermodynamic limit, but differ in a nucleus of finite nucleons. Especially in phase transition regions of sharp property changes, the differences may be large.

To cue the problems, projection onto good particle number is necessary. It is usually done by numerical integration over the gauge angle \cite{Dietrich_1964, Ring_book}, and the result is a coherent pair condensate [of generalized-seniority zero, see Eq. (\ref{gs})]. The projection can be done after or before the variation \cite{Ring_book, Bender_2003}. Projection after variation (PAV) restores the correct particle number and improves binding energy \cite{Ring_book, Anguiano_2002}. But it violates the variational principle so the energy is not at minimum. Moreover, many realistic nuclei have weak pairing and the BCS or HFB solution collapses \cite{Bertsch_2009}, then further projection gets nothing. The variation after projection (VAP) \cite{Dietrich_1964} is preferred over PAV when feasible. VAP is simply the variational principle using the coherent pair condensate as the trial wavefunction, and produces the best energy. For VAP+BCS see for example \cite{Dietrich_1964, Dussel_2007, Dukelsky_2000, Sandulescu_2008}; for VAP+HFB see for example \cite{Sheikh_2000, Anguiano_2001, Anguiano_2002, Stoitsov_2007, Hupin_2012}. The practical difficulty of VAP is that numerical projection by integration is time-consuming \cite{Wang_2014} and needed many times in the current VAP procedure. In the literature there are far fewer realistic applications of VAP than those of the HF+BCS or HFB theory without projection.


In this work we perform the variational principle directly on the coherent pair condensate (VDPC). The particle number is always conserved and the time-consuming projection is avoided. This work considers VDPC+BCS that varies the coherent pair structure $v_\alpha$ (\ref{P_dag}), and the result is equivalent to that of VAP+BCS. Our next work will extend to VDPC+HFB that varies $v_\alpha$ together with the canonical basis, and is equivalent to VDPC+HFB. (We name the new method VDPC because VAP may be misleading: there is no projection at all.)

We derive analytical expressions for the average energy and its gradient in terms of $v_\alpha$. Requiring the gradient vanishes, we get the analytical expression of $v_\alpha$ at the energy minimum, and discuss its asymptotic behavior away from (above or below) the Fermi surface. The new VDPC algorithm iterates these $v_\alpha$ expressions to minimize the energy until practically arbitrary precision. Without integration over gauge angle (necessary in the VAP formalism), the analytical expressions of this work look quite simple and allow easy physical interpretations. We demonstrate the new algorithm in a semi-realistic example using the realistic $V_{{\rm{low}}{\textrm{-}}k}$ interaction \cite{Bogner_2003} and large model spaces (up to $15$ harmonic-oscillator major shells). The energy-convergence pattern and actual computer time cost are given in detail. It is well known that zero-range pairing, frequently used together with the Skyrme density functional, diverges in the pairing channel \cite{Bulgac_2002_1, Bulgac_2002_2}; so the pairing-active space needs a phenomenological cutoff \cite{Stoitsov_2007, Wang_2014}. Our result shows that the realistic $V_{{\rm{low}}{\textrm{-}}k}$ interaction converges naturally in the pairing channel, although tiny occupation numbers (for example smaller than $10^{-5}$) contribute to the energy (by a few tens of keV).

This work relates to Refs. \cite{Dukelsky_2000, Hupin_2011, Sandulescu_2008, Hupin_2012}. The average energy of the pair condensate (\ref{gs}) has been derived in terms of $v_\alpha$, for the pairing Hamiltonian \cite{Dukelsky_2000} and a general Hamiltonian \cite{Hupin_2011}. However the gradient of energy has not been derived. Using this energy expression (without gradient) to numerically minimize energy is quick in small model spaces \cite{Sandulescu_2008, Hupin_2012}; but the numerical sign problem arises in large model spaces with many particles. For realistic applications, currently the pairing-channel model space has been limited to a $10$ MeV window around the Fermi surface \cite{Sandulescu_2008, Hupin_2012}. ($5$ MeV above and $5$ MeV below, dimension is approximately that of one major shell in atomic nuclei.) Only for the state-independent pairing Hamiltonian with uniformly spaced single-particle energies (this Hamiltonian has only one parameter of the pairing strength in unit of the single-particle energy spacing), large model spaces have been used \cite{Sandulescu_2008, Dukelsky_2000}. In this simple schematic model, $v_\alpha$ (and the occupation number) decreases monotonically as the single-particle energy increases, so probably the solution can be regulated to avoid the sign problem. Modern mean-field theories use large model spaces, and VAP has been done only by the gauge-angle integration \cite{Anguiano_2001, Anguiano_2002, Stoitsov_2007}. One aim of this work is to propose the new VDPC algorithm.

We also mention the Lipkin-Nogami prescription to restore the particle number approximately \cite{Lipkin_1960, Nogami_1964, Pradhan_1973}. It has been widely used because the exact VAP is computationally expensive \cite{Bender_2003, Wang_2014}. There are ongoing efforts to improve the Lipkin method \cite{Wang_2014}.

We also give analytical expressions for the gradient of energy with respect to changes of the canonical single-particle basis, which will be necessary for the next work in this series: VDPC+HFB.

The paper is organized as follows. Section \ref{Sec_PC} briefly reviews
the formalism for the condensate of coherent pairs. This is the ``kinematics'' of the theory. Section \ref{Sec_energy} derives the analytical expression for the average energy. Section \ref{Sec_grad_v} derives the gradient of energy with respect to $v_\alpha$, $v_\alpha$ at the energy minimum, and discusses $v_\alpha$'s asymptotic behavior away from the Fermi surface. Section \ref{Sec_grad_basis} derives the gradient of energy with respect to the canonical basis. The VDPC+BCS algorithm is described in Sec. \ref{Sec_code}, and applied to a semi-realistic example in Sec. \ref{Sec_example}. Section \ref{Sec_summary} summarizes the work.


\section{Coherent Pair Condensate  \label{Sec_PC}}

In this section we briefly review the formalism for the condensate of coherent pairs (state of zero generalized seniority \cite{Talmi_book}). For clarity we consider one kind of nucleons, the extension to active protons and neutrons is quite simple: the existence of protons simply provides a correction to the neutron single-particle energy, through the two-body proton-neutron interaction. We assume time-reversal self-consistent symmetry \cite{Ring_book, Goodman_1979}, and the single-particle basis state $|\alpha\rangle$ is Kramers degenerate with its time-reversed partner $|\tilde{\alpha}\rangle$ ($|\tilde{\tilde{\alpha}}\rangle = - |\alpha\rangle$). No other symmetry is assumed.

The ground state of the
$2N$-particle system is assumed to be an $N$-pair condensate,
\begin{eqnarray}
|\phi_N\rangle = \frac{1}{\sqrt{\chi_{N}}} (P^\dagger)^{N} |0\rangle
, \label{gs}
\end{eqnarray}
where
\begin{eqnarray}
\chi_{N} = \langle0| P^N (P^\dagger)^{N} |0\rangle  \label{chi_N}
\end{eqnarray}
is the normalization factor. The coherent pair-creation operator is
\begin{eqnarray}
P^\dagger = \frac{1}{2} \sum_{\alpha} v_{\alpha}
a_\alpha^\dagger a_{\tilde{\alpha}}^\dagger = \sum_{\alpha \in \Theta} v_{\alpha}
P^\dagger_\alpha ,  \label{P_dag}
\end{eqnarray}
where
\begin{eqnarray}
P^\dagger_\alpha = a_\alpha^\dagger a_{\tilde{\alpha}}^\dagger = P^\dagger_{\tilde{\alpha}}   \label{P_dag_alpha}
\end{eqnarray}
creates a pair on $|\alpha\rangle$ and $|\tilde{\alpha}\rangle$.
In Eq. (\ref{P_dag}), $\Theta$ is the set of pair-indices that picks only one from each degenerate pair $|\alpha\rangle$ and $|\tilde{\alpha}\rangle$. With axial symmetry, orbits of a positive magnetic quantum number are a choice for $\Theta$. $\sum_{\alpha}$ and $\sum_{\alpha \in \Theta}$ mean summing over single-particle indices and pair indices. Requiring $|\phi_N\rangle$ to be time-even implies that the pair
structure $v_{\alpha}$ (\ref{P_dag}) is real.

In practice, the canonical single-particle basis could be the self-consistent HF mean field \cite{Dussel_2007, Sandulescu_2008} or the phenomenological Nilsson level \cite{Sheikh_2000, Sheikh_2002}. In this work we vary $v_{\alpha}$ on this fixed canonical basis $|\alpha\rangle$ to minimize the average energy $\bar{E} = \langle\phi_N|H|\phi_N\rangle$. Varying the canonical basis $|\alpha\rangle$ will be in our next work of this series.

We review necessary techniques. References \cite{Jia_2015,Jia_2017} introduced the many-pair density matrix \begin{eqnarray}
t_{\alpha_1 \alpha_2 ... \alpha_p;\beta_1 \beta_2 ... \beta_q}^{[\gamma_1 \gamma_2 ... \gamma_r],M}
\equiv \langle 0 | P^{M-p} P_{\gamma_1} P_{\gamma_2} ... P_{\gamma_r}  \nonumber \\
\times P_{\alpha_1} P_{\alpha_2} ...
P_{\alpha_p}  P_{\beta_1}^\dagger P_{\beta_2}^\dagger ...
P_{\beta_q}^\dagger  \nonumber \\
\times P_{\gamma_1}^\dagger P_{\gamma_2}^\dagger ... P_{\gamma_r}^\dagger (P^\dagger)^{M-q} | 0 \rangle .  \label{t_pq}
\end{eqnarray}
All the {\emph{pair-indices}} $\alpha_1, \alpha_2, ..., \alpha_p, \beta_1, \beta_2, ..., \beta_q, \gamma_1, \gamma_2, ..., \gamma_r$ are distinct: Eq. (\ref{t_pq}) vanishes if there are duplicated $P_\mu$ operators, or duplicated $P_\mu^\dagger$ operators, owing to the Pauli
principle; and we require {\emph{by definition}} that $\alpha_1, \alpha_2, ..., \alpha_p$ and $\beta_1, \beta_2, ..., \beta_q$ have no common index (the common ones have been moved to $\gamma_1, \gamma_2, ..., \gamma_r$). $p$, $q$, $r$ are the number of $\alpha$, $\beta$, $\gamma$ indices. $M$ equals to the total number of pair-creation operators minus $r$. Physically, the $\gamma$ levels are Pauli blocked and inactive. For convenience we introduce $\{[\gamma_1 \gamma_2 ... \gamma_r]\}$ to represent a subspace of the original single-particle space, by removing Kramers pairs of single-particle levels $\gamma_1, \tilde{\gamma}_1, \gamma_2, \tilde{\gamma}_2, ... \gamma_r, \tilde{\gamma}_r$ from the latter.
Equation (\ref{t_pq}) is the pair-hopping amplitude of $P_{\alpha_1} P_{\alpha_2} ...
P_{\alpha_p}  P_{\beta_1}^\dagger P_{\beta_2}^\dagger ...
P_{\beta_q}^\dagger$ in the blocked subspace $\{[\gamma_1 \gamma_2 ... \gamma_r]\}$.

Reference \cite{Jia_2017} introduced Pauli-blocked normalizations as a special case of Eq. (\ref{t_pq}) when $p = q = 0$,
\begin{eqnarray}
\chi_{M}^{[\gamma_1 \gamma_2 ... \gamma_r]} \equiv t_{;}^{[\gamma_1 \gamma_2 ... \gamma_r],M}  \nonumber \\
= \langle0| P^M P_{\gamma_1} P_{\gamma_2} ... P_{\gamma_r} P_{\gamma_1}^\dagger P_{\gamma_2}^\dagger ... P_{\gamma_r}^\dagger (P^\dagger)^{M} |0\rangle .  \label{chi_N_blocked}
\end{eqnarray}
It is the normalization in the blocked subspace $\{[\gamma_1 \gamma_2 ... \gamma_r]\}$.
Then Ref. \cite{Jia_2017} expressed many-pair density matrix (\ref{t_pq}) by normalizations (\ref{chi_N_blocked}),
\begin{eqnarray}
t_{\alpha_1 \alpha_2 ... \alpha_p;\beta_1 \beta_2 ... \beta_q}^{[\gamma_1 \gamma_2 ... \gamma_r],M}
= \frac{(M-p)!(M-q)!}{[(M-p-q)!]^2}  \nonumber \\
\times v_{\alpha_1} v_{\alpha_2} ... v_{\alpha_p} v_{\beta_1} v_{\beta_2} ... v_{\beta_q} \chi_{M-p-q}^{[\alpha_1 \alpha_2 ... \alpha_p \beta_1 \beta_2 ... \beta_q \gamma_1 \gamma_2 ... \gamma_r]} .
  \label{t_pq_res}
\end{eqnarray}

Normalizations (\ref{chi_N_blocked}) can be computed by recursive relations,
\begin{eqnarray}
\chi_N = N \sum_{\alpha \in \Theta} (v_\alpha)^2 \chi_{N-1}^{[\alpha]} ,  \label{chi_rec1} \\
\chi_{N} - \chi_{N}^{[\alpha]} = (N v_\alpha)^2 \chi_{N-1}^{[\alpha]} = \chi_N \langle \phi_N | \hat{n}_\alpha | \phi_N \rangle ,  \label{n_ave}
\end{eqnarray}
with initial value $\chi_{N=0}^{[\alpha]} = 1$. Knowing $\chi_{N-1}^{[\alpha]}$'s, we compute $\chi_N$ by Eq. (\ref{chi_rec1}), and then $\chi_{N}^{[\alpha]}$'s by Eq. (\ref{n_ave}). $\langle \phi_N | \hat{n}_\alpha | \phi_N \rangle = \langle 0 | P^{N} a_\alpha^\dagger a_\alpha (P^\dagger)^{N} | 0 \rangle / \chi_N$ is the average occupation number. Equations (\ref{chi_rec1}) and (\ref{n_ave}) are just Eqs. (22-24) of Ref. \cite{Jia_2013_1}, using $t_{\alpha;}^N = N v_\alpha \chi_{N-1}^{[\alpha]}$ implied from Eq. (\ref{t_pq_res}).
Equations (\ref{chi_rec1}) and (\ref{n_ave}) are also valid in the blocked subspaces $\{[\gamma_1 \gamma_2 ... \gamma_r]\}$. For example, in $\{[\beta]\}$ they read
\begin{eqnarray}
\chi_N^{[\beta]} &=& N \sum_{\alpha \in \Theta}^{P_\alpha \ne P_\beta} (v_\alpha)^2 \chi_{N-1}^{[\alpha\beta]} ,  \label{chi_rec1_beta} \\
\chi_{N}^{[\beta]} - \chi_{N}^{[\alpha\beta]} &=& (N v_\alpha)^2 \chi_{N-1}^{[\alpha\beta]}  \nonumber \\
&=& \chi_N^{[\beta]} \langle \phi_N^{[\beta]} | \hat{n}_\alpha | \phi_N^{[\beta]} \rangle ~~~ (P_\alpha \ne P_\beta) ,  \label{n_ave_beta}
\end{eqnarray}
where
\begin{eqnarray}
| \phi_N^{[\beta]} \rangle \equiv \frac{1}{\sqrt{\chi_{N}^{[\beta]}}} (P^\dagger - v_\beta P_\beta^\dagger)^{N} |0\rangle
\end{eqnarray}
is the pair condensate with $\beta$ and $\tilde{\beta}$ blocked. In $\{[\beta\gamma]\}$ ($P_\beta \ne P_\gamma$) they read
\begin{eqnarray}
\chi_N^{[\beta\gamma]} &=& N \sum_{\alpha \in \Theta}^{P_\alpha \ne P_\beta,P_\gamma} (v_\alpha)^2 \chi_{N-1}^{[\alpha\beta\gamma]} ,  \label{chi_rec1_bg} \\
\chi_{N}^{[\beta\gamma]} - \chi_{N}^{[\alpha\beta\gamma]} &=& (N v_\alpha)^2 \chi_{N-1}^{[\alpha\beta\gamma]}
~~~ (P_\alpha \ne P_\beta,P_\gamma) .  \label{n_ave_bg}
\end{eqnarray}

Later the new VDPC algorithm needs $\chi_{N}^{[\alpha]}$, $\chi_{N}^{[\alpha\beta]}$, and $\chi_{N}^{[\alpha\beta\gamma]}$. $\chi_{N}^{[\alpha]}$ must be computed by the recursive relation (\ref{n_ave}). $\chi_{N}^{[\alpha\beta]}$ could be computed by the recursive relation (\ref{n_ave_beta}), but we find it simpler and quicker to use an alternative formula ($P_\alpha \ne P_\beta$):
\begin{eqnarray}
(v_\alpha)^2 \chi_{N}^{[\alpha]} - (v_\beta)^2 \chi_{N}^{[\beta]} = [(v_\alpha)^2 - (v_\beta)^2] \chi_{N}^{[\alpha\beta]} .  \label{vsq_chi}
\end{eqnarray}
Deriving Eq. (\ref{vsq_chi}) has only one step using Eq. (\ref{n_ave_beta}). Similarly, $\chi_{N}^{[\alpha\beta\gamma]}$ could be computed by the recursive relation (\ref{n_ave_bg}), but it is  simpler and quicker to use ($P_\alpha, P_\beta, P_\gamma$ are all different)
\begin{eqnarray}
(v_\alpha)^2 \chi_{N}^{[\alpha\gamma]} - (v_\beta)^2 \chi_{N}^{[\beta\gamma]}
= [(v_\alpha)^2 - (v_\beta)^2] \chi_{N}^{[\alpha\beta\gamma]} .  \label{vsq_chi_abg}
\end{eqnarray}

This section discusses the ``kinematics'' of the formalism, next we discuss the ``dynamics''.


\begin{widetext}

\section{Average Energy  \label{Sec_energy}}

In this section we derive analytically the average energy of the pair condensate. The anti-symmetrized two-body Hamiltonian is
\begin{eqnarray}
H = \sum_{\alpha} \epsilon_{\alpha\beta} a_\alpha^\dagger a_\beta + \frac{1}{4} \sum_{\alpha \beta \gamma \mu} V_{\alpha \beta \gamma \mu} a_\alpha^\dagger a_\beta^\dagger a_\gamma a_\mu . \label{H_def}
\end{eqnarray}
Note the ordering of $\alpha \beta \gamma \mu$, thus $V_{\alpha \beta \gamma \mu} = - \langle \alpha \beta| V | \gamma \mu\rangle$.
We assume time-even $H$ ($\epsilon_{\alpha\beta} = \epsilon_{\tilde{\beta}\tilde{\alpha}}$, $V_{\alpha \beta \gamma \mu} = V_{\tilde{\mu} \tilde{\gamma} \tilde{\beta} \tilde{\alpha} }$) and real $\epsilon_{\alpha\beta}$ and $V_{\alpha \beta \gamma \mu}$. No other symmetry of $H$ is assumed.

We compute the average energy $\bar{E} = \langle \phi_N | H | \phi_N \rangle$ in the canonical basis (\ref{P_dag}). For the one-body $\epsilon_{\alpha\beta}$ part, only the diagonal type $a_\alpha^\dagger a_\alpha$ contributes,
\begin{eqnarray}
\langle 0 | P^N a_\alpha^\dagger a_\alpha (P^\dagger)^N | 0 \rangle = \chi_N \langle \phi_N | \hat{n}_\alpha | \phi_N \rangle = (N v_\alpha)^2\chi_{N-1}^{[\alpha]} ,  \label{type0}
\end{eqnarray}
which is Eq. (\ref{n_ave}).
For the two-body $V_{\alpha \beta \gamma \mu}$ part, only three mutually exclusive types contribute ($P_\alpha \ne P_\beta$): $a_\alpha^\dagger a_{\tilde{\alpha}}^\dagger a_{\tilde{\alpha}} a_\alpha$, $a_\alpha^\dagger a_{\tilde{\alpha}}^\dagger a_{\tilde{\beta}} a_\beta$, and $a_\alpha^\dagger a_\beta^\dagger a_\beta a_\alpha$. The first type is
\begin{eqnarray}
type1 = \langle 0 | P^N a_\alpha^\dagger a_{\tilde{\alpha}}^\dagger a_{\tilde{\alpha}} a_\alpha (P^\dagger)^N | 0 \rangle = \langle 0 | P^N a_\alpha^\dagger a_\alpha (P^\dagger)^N | 0 \rangle = \chi_N \langle \phi_N | \hat{n}_\alpha | \phi_N \rangle ,  \label{type1}
\end{eqnarray}
because $|\alpha\rangle$ and $|\tilde{\alpha}\rangle$ are either both occupied or both empty in $(P^\dagger)^N | 0 \rangle$. The second type $a_\alpha^\dagger a_{\tilde{\alpha}}^\dagger a_{\tilde{\beta}} a_\beta = P_\alpha^\dagger P_\beta = P_\beta P_\alpha^\dagger$, so Eqs. (\ref{t_pq}) and (\ref{t_pq_res}) imply
\begin{eqnarray}
type2 = \langle 0 | P^N a_\alpha^\dagger a_{\tilde{\alpha}}^\dagger a_{\tilde{\beta}} a_\beta (P^\dagger)^N | 0 \rangle = t^{N+1}_{\beta;\alpha} = N^2 v_\alpha v_\beta \chi_{N-1}^{[\alpha\beta]} .  \label{type2}
\end{eqnarray}
The third type $a_\alpha^\dagger a_\beta^\dagger a_\beta a_\alpha = 1 - a_\alpha a_\alpha^\dagger - a_\beta a_\beta^\dagger + a_\alpha a_\beta a_\beta^\dagger a_\alpha^\dagger$ by basic anti-commutation relation, so definition (\ref{chi_N_blocked}) implies
\begin{eqnarray}
type3 = \langle 0 | P^N a_\alpha^\dagger a_\beta^\dagger a_\beta a_\alpha (P^\dagger)^N | 0 \rangle = \chi_N - \chi_N^{[\alpha]} - \chi_N^{[\beta]} + \chi_N^{[\alpha\beta]} .  \nonumber  
\end{eqnarray}
Using $\chi_N - \chi_N^{[\alpha]} = N^2 (v_\alpha)^2 \chi_{N-1}^{[\alpha]}$ [Eq. (\ref{n_ave})] and $\chi_N^{[\beta]} - \chi_N^{[\alpha\beta]} = N^2 (v_\alpha)^2 \chi_{N-1}^{[\alpha\beta]}$ [Eq. (\ref{n_ave_beta})], then factorizing out $N^2 (v_\alpha)^2$, we have
\begin{eqnarray}
type3 = N^2 (v_\alpha)^2 ( \chi_{N-1}^{[\alpha]} - \chi_{N-1}^{[\alpha\beta]} ) .  \nonumber
\end{eqnarray}
In Eq. (\ref{n_ave_beta}) we replace $N$ by $N-1$ ($N \rightarrow N-1$) and exchange $\alpha$ and $\beta$ ($\alpha \leftrightarrow \beta$),
\begin{eqnarray}
\chi_{N-1}^{[\alpha]} - \chi_{N-1}^{[\alpha\beta]} = (N-1)^2 (v_\beta)^2 \chi_{N-2}^{[\alpha\beta]} = \chi_{N-1}^{[\alpha]} \langle \phi_{N-1}^{[\alpha]} | \hat{n}_\beta | \phi_{N-1}^{[\alpha]} \rangle .  \label{n_ave_Nm1}
\end{eqnarray}
Thus we have two equivalent expressions,
\begin{eqnarray}
type3 = N^2 (N-1)^2 (v_\alpha v_\beta)^2 \chi_{N-2}^{[\alpha\beta]} ,  \label{type3_1}
\end{eqnarray}
or
\begin{eqnarray}
type3 = N^2 (v_\alpha)^2 \chi_{N-1}^{[\alpha]} \langle \phi_{N-1}^{[\alpha]} | \hat{n}_\beta | \phi_{N-1}^{[\alpha]} \rangle = \chi_{N} \langle \phi_{N} | \hat{n}_\alpha | \phi_{N} \rangle \langle \phi_{N-1}^{[\alpha]} | \hat{n}_\beta | \phi_{N-1}^{[\alpha]} \rangle  .  \label{type3_2}
\end{eqnarray}
The last step is Eq. (\ref{n_ave}).

The expectation value of $H$ (\ref{H_def}) is
\begin{eqnarray}
\langle \phi_N | H | \phi_N \rangle
= \sum_{\alpha \in \Theta} 2\epsilon_{\alpha\alpha} \langle \phi_N | a_\alpha^\dagger a_\alpha | \phi_N \rangle + \sum_{\alpha \in \Theta} V_{\alpha\tilde{\alpha}\tilde{\alpha}\alpha} \langle \phi_N | a_\alpha^\dagger a_{\tilde{\alpha}}^\dagger a_{\tilde{\alpha}} a_\alpha | \phi_N \rangle  \nonumber \\
+ \sum^{\alpha\ne\beta}_{\alpha,\beta \in \Theta} V_{\alpha\tilde{\alpha}\tilde{\beta}\beta} \langle \phi_N | a_\alpha^\dagger a_{\tilde{\alpha}}^\dagger a_{\tilde{\beta}} a_\beta | \phi_N \rangle + \sum^{\alpha \ne \beta}_{\alpha,\beta \in \Theta} (V_{\alpha\beta\beta\alpha} + V_{\alpha\tilde{\beta}\tilde{\beta}\alpha}) \langle \phi_N | a_\alpha^\dagger a_{\beta}^\dagger a_\beta a_\alpha | \phi_N \rangle .  \label{H_der1}
\end{eqnarray}
The summations run over pair-index $\alpha$ and $\beta$.
The first term collects two equal contributions by $\epsilon_{\alpha\alpha} = \epsilon_{\tilde{\alpha}\tilde{\alpha}}$, which gives the factor $2$.
The second term collects four equal contributions by $V_{\alpha\tilde{\alpha}\tilde{\alpha}\alpha} = - V_{\alpha\tilde{\alpha}\alpha\tilde{\alpha}} = - V_{\tilde{\alpha}\alpha\tilde{\alpha}\alpha} = V_{\tilde{\alpha}\alpha\alpha\tilde{\alpha}}$, which cancels the factor $1/4$ in the Hamiltonian (\ref{H_def}). The third term collects $V_{\alpha\tilde{\alpha}\tilde{\beta}\beta} = - V_{\alpha\tilde{\alpha}\beta\tilde{\beta}} = - V_{\tilde{\alpha}\alpha\tilde{\beta}\beta} = V_{\tilde{\alpha}\alpha\beta\tilde{\beta}}$. The fourth term collects $V_{\alpha\beta\beta\alpha} = - V_{\alpha\beta\alpha \beta} = V_{\tilde{\alpha}\tilde{\beta}\tilde{\beta}\tilde{\alpha}} = - V_{\tilde{\alpha}\tilde{\beta}\tilde{\alpha}\tilde{\beta}}$ and $V_{\alpha\tilde{\beta}\tilde{\beta}\alpha} = - V_{\alpha\tilde{\beta}\alpha \tilde{\beta}} = V_{\tilde{\alpha}\beta\beta\tilde{\alpha}} = - V_{\tilde{\alpha}\beta\tilde{\alpha}\beta}$.

Substituting Eqs. (\ref{type0},\ref{type1},\ref{type2},\ref{type3_1}) into Eq. (\ref{H_der1}),
\begin{eqnarray}
\langle \phi_N | H | \phi_N \rangle = \frac{N^2}{\chi_N} \Big( \sum_{\alpha \in \Theta} (2\epsilon_{\alpha\alpha} + G_{\alpha\alpha}) (v_\alpha)^2 \chi_{N-1}^{[\alpha]}
+ \sum^{\alpha \ne \beta}_{\alpha,\beta \in \Theta} G_{\alpha\beta} v_\alpha v_\beta \chi_{N-1}^{[\alpha\beta]} + (N-1)^2 \sum^{\alpha\ne\beta}_{\alpha,\beta \in \Theta} \Lambda_{\alpha\beta} (v_\alpha v_\beta)^2 \chi_{N-2}^{[\alpha\beta]} \Big) ,  \label{H_ave_1}
\end{eqnarray}
where we introduce the paring matrix elements $G_{\alpha\beta}$ and the ``monopole'' matrix elements $\Lambda_{\alpha\beta}$ as
\begin{eqnarray}
G_{\alpha\beta} \equiv V_{\alpha\tilde{\alpha}\tilde{\beta}\beta} ,  \label{G_12} \\
\Lambda_{\alpha\beta} = V_{\alpha\beta\beta\alpha} + V_{\alpha\tilde{\beta}\tilde{\beta}\alpha} . \label{L_12}
\end{eqnarray}
Note $G_{\alpha\beta} = G_{\beta\alpha} = G_{\alpha\tilde{\beta}}$, $\Lambda_{\alpha\beta} = \Lambda_{\beta\alpha} = \Lambda_{\alpha\tilde{\beta}}$, and $G_{\alpha\alpha} = \Lambda_{\alpha\alpha}$. Equation (\ref{H_ave_1}) expresses the average energy by normalizations and is adopted in coding. Another equivalent expression by occupation numbers reveals more physics. Using Eqs. (\ref{type0}) and (\ref{type3_2}),
\begin{eqnarray}
\langle \phi_N | H | \phi_N \rangle = \sum_{\alpha \in \Theta} (2\epsilon_{\alpha\alpha} + G_{\alpha\alpha}) \langle \phi_N | \hat{n}_\alpha | \phi_N \rangle
+ N^2 \sum^{\alpha \ne \beta}_{\alpha,\beta \in \Theta} G_{\alpha\beta} v_\alpha v_\beta \frac{\chi_{N-1}^{[\alpha\beta]}}{\chi_N} + \sum^{\alpha\ne\beta}_{\alpha,\beta \in \Theta} \Lambda_{\alpha\beta} \langle \phi_N | \hat{n}_\alpha | \phi_N \rangle \langle \phi_{N-1}^{[\alpha]} | \hat{n}_\beta | \phi_{N-1}^{[\alpha]} \rangle .~~  \label{H_ave_2}
\end{eqnarray}

The analytical expression for the average energy has been given in a slightly different form in Ref. \cite{Hupin_2011}. The gradient of energy and others, crucial to the new VDPC algorithm, are new results of this work as given in the next section.


\section{Gradient of Energy with respect to Pair Structure  \label{Sec_grad_v}}

In this section we derive the gradient of average energy with respect to the pair structure $v_\alpha$ (\ref{P_dag}). Moreover, we give the analytical formula of $v_\alpha$ that minimizes the energy. Finally we discuss the asymptotic behavior of $v_\alpha$ away from (above or below) the Fermi surface.

Equation (\ref{H_ave_1}) expresses the average energy $\bar{E}$ in terms of (Pauli-blocked) normalizations $\chi_N$. To compute gradient of $\bar{E}$, we first compute gradient of $\chi_N$. Under infinitesimal change $\delta v_\alpha$ of a single $v_\alpha$, the variation of $P^\dagger$ (\ref{P_dag}) and $(P^\dagger)^N$ are
\begin{eqnarray}
\delta P^\dagger = (\delta v_{\alpha}) P^\dagger_\alpha ,  \nonumber \\
\delta [(P^\dagger)^N] = N (P^\dagger)^{N-1} \delta P^\dagger = N (\delta v_{\alpha}) P^\dagger_\alpha (P^\dagger)^{N-1} .  \nonumber
\end{eqnarray}
Thus the variation of $\chi_N$ (\ref{chi_N}) is
\begin{eqnarray}
\delta \chi_{N} = \langle0| \delta[P^N] (P^\dagger)^{N} |0\rangle + \langle0| P^N \delta[(P^\dagger)^{N}] |0\rangle = \langle0| P^N \delta[(P^\dagger)^{N}] |0\rangle + h.c. = N \langle0| P^N P^\dagger_\alpha (P^\dagger)^{N-1}
|0\rangle \delta v_{\alpha} + h.c.   \nonumber
\end{eqnarray}
h.c. means hermitian conjugate, which in fact equals to the first term because $v_\alpha$ is real. Using Eqs. (\ref{t_pq}) and (\ref{t_pq_res}), we have
\begin{eqnarray}
\delta \chi_{N} = 2 N t_{;\alpha}^N \delta v_{\alpha}
= 2 N^2 v_\alpha \chi_{N-1}^{[\alpha]} \delta v_{\alpha} = \frac{2 \chi_N}{v_\alpha} \langle \phi_N | \hat{n}_\alpha | \phi_N \rangle \delta v_{\alpha} = \frac{2}{v_\alpha} (\chi_{N} - \chi_{N}^{[\alpha]}) \delta v_{\alpha} .  \label{delta_chi}
\end{eqnarray}
The last two steps use Eq. (\ref{n_ave}).

If we Pauli block the $\beta$ index ($P_\beta \ne P_\alpha$) from the very beginning, the derivation remains valid, so Eq. (\ref{delta_chi}) implies
\begin{eqnarray}
\delta \chi_{N}^{[\beta]} = 2 N^2 v_\alpha \chi_{N-1}^{[\alpha\beta]} \delta v_{\alpha} = \frac{2 \chi_N^{[\beta]}}{v_\alpha} \langle \phi_N^{[\beta]} | \hat{n}_\alpha | \phi_N^{[\beta]} \rangle \delta v_{\alpha} = \frac{2}{v_\alpha} (\chi_{N}^{[\beta]} - \chi_{N}^{[\alpha\beta]}) \delta v_{\alpha} ,~~~ P_\beta \ne P_\alpha .  \label{delta_chibeta}
\end{eqnarray}
And of cause $\delta \chi_{N}^{[\alpha]} = 0$. Similarly, we easily obtain $\delta \chi_{N}^{[\beta\gamma]}$ by Pauli blocking the two indices $\beta$ and $\gamma$ from the very beginning. Substituting $\delta \chi_{N}$ (\ref{delta_chi}), $\delta \chi_{N}^{[\beta]}$ (\ref{delta_chibeta}), and $\delta \chi_{N}^{[\beta\gamma]}$ into Eq. (\ref{H_ave_1}), using basic calculus then collecting similar terms, a two-page long derivation gives the energy gradient,
\begin{eqnarray}
\frac{\partial \bar{E}}{\partial v_{\alpha}} = \frac{\partial[ \langle \phi_N | H | \phi_N \rangle ]}{\partial v_{\alpha}}
=^{exp1~} \frac{- 2N^2 }{\chi_{N}} \Big[ \sum^{P_\beta \ne P_\alpha}_{\beta \in \Theta} G_{\alpha\beta} v_\beta \chi_{N-1}^{[\alpha\beta]} + \frac{\chi_{N}^{[\alpha]}}{N^2 v_\alpha} (\langle \phi_N^{[\alpha]} | H | \phi_N^{[\alpha]} \rangle - \bar{E}) \Big]  \label{grad_v_1} \\
=^{exp2~} \frac{2 N^2}{\chi_N} \Big[ \sum^{P_\beta \ne P_\alpha}_{\beta \in \Theta} G_{\alpha\beta} v_\beta \chi_{N-1}^{[\alpha\beta]}
+ v_\alpha \chi_{N-1}^{[\alpha]} ( d_{\alpha} + \langle \phi_{N-1}^{[\alpha]} | H | \phi_{N-1}^{[\alpha]} \rangle - \bar{E} ) \Big] ,  \label{grad_v_2}
\end{eqnarray}
where
\begin{eqnarray}
d_{\alpha} = 2 \epsilon_{\alpha\alpha} + G_{\alpha\alpha} + 2 \sum^{P_\beta \ne P_\alpha}_{\beta \in \Theta} \Lambda_{\alpha\beta} \langle \phi_{N-1}^{[\alpha]} | \hat{n}_\beta | \phi_{N-1}^{[\alpha]} \rangle  \label{d_a_1} \\
= 2 \epsilon_{\alpha\alpha} + G_{\alpha\alpha} + 2 (N-1)^2 \sum^{P_\beta \ne P_\alpha}_{\beta \in \Theta} \Lambda_{\alpha\beta} (v_\beta)^2 \frac{\chi_{N-2}^{[\alpha\beta]}}{\chi_{N-1}^{[\alpha]}} .  \label{d_a_2}
\end{eqnarray}
The two gradient expressions (\ref{grad_v_1}) and (\ref{grad_v_2}) are equivalent. $d_\alpha$ is the single-pair energy similar to the common single-particle HF energy. In Eq. (\ref{d_a_1}), $2 \epsilon_{\alpha\alpha} + G_{\alpha\alpha}$ is the unperturbed energy for a pair on orbits $|\alpha\rangle$ and $|\tilde{\alpha}\rangle$, this pair interacts with other pairs by energy $2 \sum^{P_\beta \ne P_\alpha}_{\beta \in \Theta} \Lambda_{\alpha\beta} \langle \phi_{N-1}^{[\alpha]} | \hat{n}_\beta | \phi_{N-1}^{[\alpha]} \rangle$. Equation (\ref{d_a_2}) is equivalent to Eq. (\ref{d_a_1}), based on Eq. (\ref{n_ave_beta}).


Because $\bar{E}$ is independent of an overall scaling of $v_\alpha$, the gradient of $\bar{E}$ is perpendicular to $\vec{v}$,
\begin{eqnarray}
\nabla \bar{E} \cdot \vec{v} = \sum_{\alpha \in \Theta} v_\alpha \frac{\partial \bar{E}}{\partial v_{\alpha}} = 0 .  \nonumber
\end{eqnarray}
This identity is used to numerically check the computer code.

Later we also need the HF single-particle energy
\begin{eqnarray}
e_\alpha = \epsilon_{\alpha\alpha} + \sum_{\beta \in {\rm{SD}}} V_{\alpha\beta\beta\alpha} = \epsilon_{\alpha\alpha} + \sum_{\beta \in \Theta}^{\beta \in {\rm{SD}}} \Lambda_{\alpha\beta} ,   \label{e_HF}
\end{eqnarray}
where $\beta \in {\rm{SD}}$ means the $\beta$ orbit is occupied in the HF Slater determinant. In Eq. (\ref{P_dag}), if we set $v_\alpha$ to $1$ for occupied orbits and to $0$ for empty orbits, the pair condensate (\ref{gs}) reduces to the HF Slater determinant. In this case $d_\alpha \approx 2 e_\alpha$.


At energy minimum, the gradient (\ref{grad_v_1}) and (\ref{grad_v_2}) vanish, which implies
\begin{eqnarray}
v_\alpha =^{exp1~} \frac{\langle \phi_N^{[\alpha]} | H | \phi_N^{[\alpha]} \rangle - \bar{E}}{- N^2 ( \sum\limits^{P_\beta \ne P_\alpha}_{\beta \in \Theta} G_{\alpha\beta} v_\beta \chi_{N-1}^{[\alpha\beta]} ) / \chi_{N}^{[\alpha]}} = \frac{\langle \phi_N^{[\alpha]} | H | \phi_N^{[\alpha]} \rangle - \bar{E}}{- \sum\limits^{P_\beta \ne P_\alpha}_{\beta \in \Theta} G_{\alpha\beta} \frac{1}{v_\beta} \langle \phi_N^{[\alpha]} | {\hat{n}}_{\beta} | \phi_N^{[\alpha]} \rangle }  \label{v_1}  \\
=^{exp2~}  \frac{ - (\sum\limits^{P_\beta \ne P_\alpha}_{\beta \in \Theta} G_{\alpha\beta} v_\beta \chi_{N-1}^{[\alpha\beta]}) / \chi_{N-1}^{[\alpha]} }{ d_{\alpha} + \langle \phi_{N-1}^{[\alpha]} | H | \phi_{N-1}^{[\alpha]} \rangle - \bar{E} } = \frac{ - \sum\limits^{P_\beta \ne P_\alpha}_{\beta \in \Theta} G_{\alpha\beta} v_\beta (1-\langle \phi_{N-1}^{[\alpha]} | {\hat{n}}_{\beta} | \phi_{N-1}^{[\alpha]} \rangle)}{ d_{\alpha} + \langle \phi_{N-1}^{[\alpha]} | H | \phi_{N-1}^{[\alpha]} \rangle - \bar{E} } .  \label{v_2}
\end{eqnarray}
The last step of Eqs. (\ref{v_1}) and (\ref{v_2}) use Eq. (\ref{n_ave_beta}). Analytically, Eq. (\ref{grad_v_1}) is equivalent to Eq. (\ref{grad_v_2}), and Eq. (\ref{v_1}) is equivalent to Eq. (\ref{v_2}). Numerically, we prefer Eq. (\ref{grad_v_1}) or Eq. (\ref{v_1}) when $e_\alpha \ll e_F$ ($e_F$ is the Fermi energy, here ``$\ll$'' means the $\alpha$ orbit is well below the Fermi surface), and prefer Eq. (\ref{grad_v_2}) or Eq. (\ref{v_2}) when $e_\alpha \gg e_F$. When $e_\alpha \ll e_F$, physical arguments imply $\langle \phi_N^{[\alpha]} | H | \phi_N^{[\alpha]} \rangle - \bar{E} \approx 2 (e_F-e_\alpha)$ and $d_{\alpha} + \langle \phi_{N-1}^{[\alpha]} | H | \phi_{N-1}^{[\alpha]} \rangle - \bar{E} \approx 0$. We want to avoid the ``$\approx 0$'' case to avoid the numerical sign problem (subtract two very closed numbers, $d_{\alpha} + \langle \phi_{N-1}^{[\alpha]} | H | \phi_{N-1}^{[\alpha]} \rangle$ and $\bar{E}$), so we prefer Eq. (\ref{grad_v_1}) or Eq. (\ref{v_1}). When $e_\alpha \gg e_F$, physical arguments imply $\langle \phi_N^{[\alpha]} | H | \phi_N^{[\alpha]} \rangle - \bar{E} \approx 0$ and $d_{\alpha} + \langle \phi_{N-1}^{[\alpha]} | H | \phi_{N-1}^{[\alpha]} \rangle - \bar{E} \approx 2 (e_\alpha-e_F)$. We want to avoid the $\langle \phi_N^{[\alpha]} | H | \phi_N^{[\alpha]} \rangle - \bar{E} \approx 0$ case, so prefer Eq. (\ref{grad_v_2}) or Eq. (\ref{v_2}).

The above analysis also implies the asymptotic behavior of $v_\alpha$ away from (above or below) the Fermi surface,
\begin{eqnarray}
v_{\alpha} \approx \frac{2(e_F - e_\alpha)}{- N^2 ( \sum\limits^{P_\beta \ne P_\alpha}_{\beta \in \Theta} G_{\alpha\beta} v_\beta \chi_{N-1}^{[\alpha\beta]} ) / \chi_{N}^{[\alpha]} } = \frac{2(e_F - e_\alpha)}{- \sum\limits^{P_\beta \ne P_\alpha}_{\beta \in \Theta} G_{\alpha\beta} \frac{1}{v_\beta} \langle \phi_N^{[\alpha]} | {\hat{n}}_{\beta} | \phi_N^{[\alpha]} \rangle} ~,~~~ e_\alpha \ll e_F ,  \label{v_L}
\end{eqnarray}
and
\begin{eqnarray}
v_\alpha \approx \frac{ - ( \sum\limits^{P_\beta \ne P_\alpha}_{\beta \in \Theta} G_{\alpha\beta} v_\beta \chi_{N-1}^{[\alpha\beta]} ) / \chi_{N-1}^{[\alpha]} }{2(e_\alpha - e_F)} = \frac{ - \sum\limits^{P_\beta \ne P_\alpha}_{\beta \in \Theta} G_{\alpha\beta} v_\beta (1-\langle \phi_{N-1}^{[\alpha]} | {\hat{n}}_{\beta} | \phi_{N-1}^{[\alpha]} \rangle) }{2(e_\alpha - e_F)} ~,~~~ e_\alpha \gg e_F .  \label{v_H}
\end{eqnarray}
In Eqs. (\ref{v_L}) and (\ref{v_H}), the summation
\begin{eqnarray}
\sum\limits^{P_\beta \ne P_\alpha}_{\beta \in \Theta} G_{\alpha\beta} v_\beta \chi_{N-1}^{[\alpha\beta]} = \frac{\chi_{N}^{[\alpha]}}{N^2} \sum\limits^{P_\beta \ne P_\alpha}_{\beta \in \Theta} G_{\alpha\beta} \frac{1}{v_\beta} \langle \phi_N^{[\alpha]} | {\hat{n}}_{\beta} | \phi_N^{[\alpha]} \rangle  \label{Gv_sum_1} \\
= \chi_{N-1}^{[\alpha]} \sum\limits^{P_\beta \ne P_\alpha}_{\beta \in \Theta} G_{\alpha\beta} v_\beta (1-\langle \phi_{N-1}^{[\alpha]} | {\hat{n}}_{\beta} | \phi_{N-1}^{[\alpha]} \rangle)  \label{Gv_sum_2}
\end{eqnarray}
depends on the details of interaction, and should have important contributions when the $\beta$ orbit is near the Fermi surface. If $e_\beta \ll e_F$, $v_\beta$ is very large and $\langle \phi_N^{[\alpha]} | {\hat{n}}_{\beta} | \phi_N^{[\alpha]} \rangle \approx 1$, Eq. (\ref{Gv_sum_1}) shows that this $G_{\alpha\beta}$ term is suppressed by the factor $1/v_\beta$, compared with those terms near the Fermi surface. If $e_\beta \gg e_F$, $v_\beta$ is very small and $\langle \phi_{N-1}^{[\alpha]} | {\hat{n}}_{\beta} | \phi_{N-1}^{[\alpha]} \rangle \approx 0$, Eq. (\ref{Gv_sum_2}) shows that this $G_{\alpha\beta}$ term is suppressed by the factor $v_\beta$, compared with those terms near the Fermi surface.

When $e_\alpha \ll e_F$, generally $v_\alpha$ should increase when $e_\alpha$ decreases, and the linear factor $e_F - e_\alpha$ in numerator of Eq. (\ref{v_L}) contributes to this trend. The other factor $\sum\limits^{P_\beta \ne P_\alpha}_{\beta \in \Theta} G_{\alpha\beta} \frac{1}{v_\beta} \langle \phi_N^{[\alpha]} | {\hat{n}}_{\beta} | \phi_N^{[\alpha]} \rangle$ in denominator should also contribute to this trend by the decaying of $G_{\alpha\beta}$, because $\beta$ was near the Fermi surface as explained above. When $e_\alpha \gg e_F$, generally $v_\alpha$ should decrease when $e_\alpha$ increases, and the inverse-linear factor $1/(e_\alpha - e_F)$ in denominator of Eq. (\ref{v_H}) contributes to this trend. The other factor $\sum\limits^{P_\beta \ne P_\alpha}_{\beta \in \Theta} G_{\alpha\beta} v_\beta (1-\langle \phi_{N-1}^{[\alpha]} | {\hat{n}}_{\beta} | \phi_{N-1}^{[\alpha]} \rangle)$ in numerator should also contribute to this trend by the decaying of $G_{\alpha\beta}$.

The exact [Eqs. (\ref{v_1}) and (\ref{v_2})] and asymptotic [Eqs. (\ref{v_L}) and (\ref{v_H})] expressions for $v_\alpha$ are key to the new VDPC algorithm as given in Sec. \ref{Sec_code}.



\section{Gradient of Energy with respect to Canonical Basis  \label{Sec_grad_basis}}

In VDPC + HFB, the two sets of variational parameters are the unitary transformation to the canonical single-particle basis, and the pair structure $v_\alpha$ (\ref{P_dag}) on the canonical basis. In Sec. \ref{Sec_grad_v} we have derived the gradient of energy with respect to $v_\alpha$, which is enough for VDPC + BCS. In this section we derive analytically the gradient of energy with respect to changes of the canonical basis, and transform the energy minimization (varying the canonical basis while fixing $v_\alpha$) into a diagonalization problem.

We parameterize the mixing of two single-particle levels as
\begin{eqnarray}
|\alpha'\rangle = \cos \theta |\alpha\rangle + \sin \theta |2\rangle ~,~ |\beta'\rangle = \cos \theta |\beta\rangle - \sin \theta |\alpha\rangle ,  \label{tran_ori}
\end{eqnarray}
consequently for their time-reversal partners
\begin{eqnarray}
|\widetilde{\alpha'}\rangle = \cos \theta |\tilde{\alpha}\rangle + \sin \theta |\tilde{\beta}\rangle ~,~ |\widetilde{\beta'}\rangle = \cos \theta |\tilde{\beta}\rangle - \sin \theta |\tilde{\alpha}\rangle .  \label{tran_time}
\end{eqnarray}
$|\alpha\rangle$ and $|\beta\rangle$ belong to different Kramers pairs ($P_\alpha \ne P_{\beta}$): the pair condensate $|\phi_N\rangle$ (\ref{gs}) is invariant under mixing of $|\alpha\rangle$ and $|\tilde{\alpha}\rangle$. (Because $P_\alpha^\dagger = a_\alpha^\dagger a_{\tilde{\alpha}}^\dagger$ (\ref{P_dag_alpha}) is invariant.) When there is additional symmetry (for example axial symmetry), $|\alpha\rangle$ and $|\beta\rangle$ have the same values of good quantum numbers (angular momentum projection onto the symmetry axis and parity). For infinitesimal transformation, keeping first-order in $\theta$, Eqs. (\ref{tran_ori}) and (\ref{tran_time}) imply
\begin{eqnarray}
\delta |\alpha\rangle = |\alpha'\rangle - |\alpha\rangle \approx \theta |\beta\rangle ~,~ \delta |\beta\rangle = |\beta'\rangle - |\beta\rangle \approx - \theta |\alpha\rangle ,  \label{d_ori} \\
\delta |\tilde{\alpha}\rangle = |\widetilde{\alpha'}\rangle - |\tilde{\alpha}\rangle \approx \theta |\tilde{\beta}\rangle ~,~ \delta |\tilde{\beta}\rangle = |\widetilde{\beta'}\rangle - |\tilde{\beta}\rangle \approx - \theta |\tilde{\alpha}\rangle .  \label{d_time}
\end{eqnarray}
We vary the canonical basis through replacing $|\alpha\rangle$, $|\tilde{\alpha}\rangle$, $|\beta\rangle$, $|\tilde{\beta}\rangle$ by $|\alpha'\rangle$, $|\widetilde{\alpha'}\rangle$, $|\beta'\rangle$, $|\widetilde{\beta'}\rangle$, and keeping other orbits unchanged. The pair condensate $|\phi_N\rangle$ (\ref{gs}) changes to $|\phi_N'\rangle$, and the average energy $\bar{E} = \langle \phi_N | H | \phi_N \rangle$ (\ref{H_ave_1}) changes to $\bar{E}' = \langle \phi_N' | H | \phi_N' \rangle$. There are two ways to compute $\bar{E}'$, by writing $H$ and $| \phi_N' \rangle$ in the new canonical basis $|\alpha'\rangle$, or by writing them in the old canonical basis $|\alpha\rangle$. We checked the two ways give the same results. The first way is easier and we show it here. In the new basis $|\alpha'\rangle$, the numerical values of $v_{\alpha'} = v_\alpha$ stay unchanged. Consequently in Eq. (\ref{H_ave_1}) the numerical values of $\chi_N$, $\chi_{N-1}^{[\alpha]}$, $\chi_{N-1}^{[\alpha\beta]}$, and $\chi_{N-2}^{[\alpha\beta]}$ stay unchanged. But $\epsilon_{\alpha\alpha}$, $G_{\alpha\alpha}$, $G_{\alpha\beta}$, $\Lambda_{\alpha\beta}$ change to their matrix representation in the new basis $|\alpha'\rangle$. Keeping first-order in $\theta$, the variations of $\epsilon_{\alpha\beta} = \langle \alpha | \epsilon | \beta \rangle$ and $V_{\alpha\beta\gamma\mu} = - \langle \alpha\beta | V | \gamma\mu \rangle$ (\ref{H_def}) are
\begin{eqnarray}
\delta \epsilon_{\alpha\beta} = \epsilon_{\alpha'\beta'} - \epsilon_{\alpha\beta} = \epsilon_{\delta\alpha,\beta} + \epsilon_{\alpha,\delta\beta} ,  \label{var_epsilon}  \\
\delta V_{\alpha\beta\gamma\mu} = V_{\alpha'\beta'\gamma'\mu'} - V_{\alpha\beta\gamma\mu} = V_{\delta\alpha,\beta,\gamma,\mu} + V_{\alpha,\delta\beta,\gamma,\mu} + V_{\alpha,\beta,\delta\gamma,\mu} + V_{\alpha,\beta,\gamma,\delta\mu} .  \label{var_V}
\end{eqnarray}
In Eqs. (\ref{var_epsilon}) and (\ref{var_V}), $\alpha$ and $\beta$ are generic and not restricted to the two orbits being mixed (\ref{d_ori}). These two equations are used to compute variations of $\epsilon_{\alpha\alpha}$, $G_{\alpha\alpha}$, $G_{\alpha\beta}$, $\Lambda_{\alpha\beta}$ in Eq. (\ref{H_ave_1}). We compute $G_{\alpha\beta}$ (\ref{G_12}) as an example,
\begin{eqnarray}
\delta G_{\alpha\beta} = \delta V_{\alpha\tilde{\alpha}\tilde{\beta}\beta} = V_{\delta \alpha,\tilde{\alpha},\tilde{\beta},\beta} + V_{\alpha,\delta\tilde{\alpha},\tilde{\beta},\beta} + V_{\alpha,\tilde{\alpha},\delta \tilde{\beta},\beta} + V_{\alpha,\tilde{\alpha},\tilde{\beta},\delta \beta}  \nonumber \\
= \theta V_{\beta\tilde{\alpha}\tilde{\beta}\beta} + \theta V_{\alpha\tilde{\beta}\tilde{\beta}\beta} - \theta V_{\alpha\tilde{\alpha}\tilde{\alpha}\beta} - \theta V_{\alpha\tilde{\alpha}\tilde{\beta}\alpha} = 2 \theta (V_{\alpha\tilde{\beta}\tilde{\beta}\beta} - V_{\alpha\tilde{\beta}\tilde{\alpha}\alpha}) .  \nonumber
\end{eqnarray}
The second last step uses Eqs. (\ref{d_ori}) and (\ref{d_time}). For other Hamiltonian parameters in Eq. (\ref{H_ave_1}), the results are ($P_\gamma \ne P_\alpha, P_\beta$)
\begin{eqnarray}
\delta \epsilon_{\alpha\alpha} = - \delta \epsilon_{\beta\beta} = 2 \theta \epsilon_{\alpha\beta} ,  \label{d_epsilon_11}  \\
\delta G_{\alpha\alpha} = 4 \theta V_{\alpha\tilde{\beta} \tilde{\alpha}\alpha} ,~
\delta G_{\beta\beta} = - 4 \theta V_{\alpha\tilde{\beta}\tilde{\beta}\beta} ,  \label{d_G_11}  \\
\delta G_{\alpha\beta} = 2 \theta (V_{\alpha\tilde{\beta}\tilde{\beta}\beta} - V_{\alpha\tilde{\beta}\tilde{\alpha}\alpha}) ~,~
\delta G_{\alpha\gamma} =
- \delta G_{\gamma\beta} = 2 \theta V_{\alpha\tilde{\beta}\tilde{\gamma}\gamma} ,  \label{d_G_12}  \\
\delta \Lambda_{\alpha\beta} = 2 \theta ( V_{\alpha\tilde{\beta}\tilde{\beta}\beta} - V_{\alpha\tilde{\beta}\tilde{\alpha}\alpha} ) ~,~
\delta \Lambda_{\alpha\gamma} = - \delta \Lambda_{\gamma\beta} = 2 \theta ( V_{\alpha\gamma\gamma\beta} + V_{\alpha\tilde{\gamma}\tilde{\gamma}\beta} ) .  \label{d_L_12}
\end{eqnarray}
Substituting them into Eq. (\ref{H_ave_1}), a one-page long derivation gives
\begin{eqnarray}
\delta \bar{E} = \delta (\langle \phi_N | H | \phi_N \rangle) = 4 \theta f_{\alpha\beta} ,  \label{E_var}
\end{eqnarray}
where
\begin{eqnarray}
f_{\alpha\beta} \equiv \frac{N^2 (v_\alpha - v_\beta)}{\chi_N } \Big( [ (v_\alpha + v_\beta) \epsilon_{\alpha\beta} + v_\alpha V_{\alpha\tilde{\beta}\tilde{\alpha}\alpha}
+ v_\beta V_{\alpha\tilde{\beta}\tilde{\beta}\beta} ] \chi_{N-1}^{[\alpha\beta]}  \nonumber \\
+ \sum_{\gamma \in \Theta}^{P_\gamma \ne P_\alpha, P_\beta} v_\gamma V_{\alpha\tilde{\beta}\tilde{\gamma}\gamma} [ \chi_{N-1}^{[\alpha\beta\gamma]} - (N-1)^2 v_\alpha v_\beta \chi_{N-2}^{[\alpha\beta\gamma]}]
+ (N-1)^2 (v_\alpha + v_\beta) \sum_{\gamma \in \Theta}^{P_\gamma \ne P_\alpha, P_\beta} (v_\gamma)^2 ( V_{\alpha\gamma\gamma\beta} + V_{\alpha\tilde{\gamma}\tilde{\gamma}\beta} ) \chi_{N-2}^{[\alpha\beta\gamma]} \Big) .  \label{f_12}
\end{eqnarray}
$f_{\alpha\beta}$ is a real anti-symmetric
matrix. We pull out the factor $4$ in Eq. (\ref{E_var}), so that $f_{\alpha\beta}$ reduces to the off-diagonal part of the HF mean field when the pair condensate (\ref{gs}) reduces to a Slater determinant. Using $(N-1)^2 (v_\gamma)^2 \chi_{N-2}^{[\alpha\beta\gamma]} = \chi_{N-1}^{[\alpha\beta]} \langle \phi_{N-1}^{[\alpha\beta]} | \hat{n}_\gamma | \phi_{N-1}^{[\alpha\beta]} \rangle$
[Eq. (\ref{n_ave}) with $N \rightarrow N-1$, $\alpha \rightarrow \gamma$, then blocking $\alpha$ and $\beta$ Kramers pairs], an equivalent form of $f_{\alpha\beta}$ is
\begin{eqnarray}
f_{\alpha\beta} = \frac{N^2 (v_\alpha - v_\beta)}{\chi_N } \Big( [ (v_\alpha + v_\beta) (\epsilon_{\alpha\beta} + \sum_{\gamma}^{P_\gamma \ne P_\alpha, P_\beta} V_{\alpha\gamma\gamma\beta} \langle \phi_{N-1}^{[\alpha\beta]} | \hat{n}_\gamma | \phi_{N-1}^{[\alpha\beta]} \rangle ) + v_\alpha V_{\alpha\tilde{\beta}\tilde{\alpha}\alpha}
+ v_\beta V_{\alpha\tilde{\beta}\tilde{\beta}\beta} ] \chi_{N-1}^{[\alpha\beta]}  \nonumber \\
+ \frac{1}{2} \sum_{\gamma}^{P_\gamma \ne P_\alpha, P_\beta} v_\gamma V_{\alpha\tilde{\beta}\tilde{\gamma}\gamma} [ \chi_{N-1}^{[\alpha\beta\gamma]} - (N-1)^2 v_\alpha v_\beta \chi_{N-2}^{[\alpha\beta\gamma]}]
 \Big) ,  \label{f_12_2}
\end{eqnarray}
where $\gamma$ sums over single-particle index.

The diagonal matrix elements $f_{\alpha\alpha} \sim (v_\alpha - v_\alpha)$ vanish. We define the full self-consistent mean field of the pair condensate as
\begin{eqnarray}
h_{\alpha\beta} \equiv  \left\{
\begin{array}{c}
{\rm{sgn}}(e_\beta - e_\alpha) ~ f_{\alpha\beta} ~,~~~ \alpha \ne \beta \\
d_\alpha/2 ~,~~~ \alpha = \beta
\end{array}  \right.
.  \label{h_12}
\end{eqnarray}
$d_\alpha$ is the single-pair energy defined in Eqs. (\ref{d_a_1}) and (\ref{d_a_2}). sgn$(e_\beta - e_\alpha)$ is the sign function that returns $1$ when $e_\beta \ge e_\alpha$ and $-1$ when $e_\beta < e_\alpha$.
$h_{\alpha\beta}$ is a real symmetric (Hermitian) matrix. At energy minimum $\delta \bar{E} = 0$, so $f_{\alpha\beta}$ vanishes and $h_{\alpha\beta}$ is diagonal. Practically, the VDPC algorithm iteratively diagonalizes $h_{\alpha\beta}$ to minimize energy with respect to the canonical basis (similarly to solving the HF equation).

The current work considers VDPC + BCS, so this section is actually not needed. This section is necessary for our next work in the series: VDPC + HFB.

%

\end{widetext}

\section{Computer Algorithm  \label{Sec_code}}

In this section we design the computer algorithm to minimize the average energy $\bar{E}$. The variational parameters are the pair structure $v_\alpha$ (\ref{P_dag}). We have expressed $\bar{E}$ (\ref{H_ave_1}) and its gradient $\frac{\partial \bar{E}}{\partial v_{\alpha}}$ [Eqs. (\ref{grad_v_1}) and (\ref{grad_v_2})] in terms of $v_\alpha$. In principle, coding these expressions and choosing an available minimizer (for example {\emph{fminunc}} in matlab) solve the problem.

Practically, in large model spaces the sign problem may arise. If we normalize $v_\alpha$ such that $v_\alpha \approx 1$ (the order of magnitude) near the Fermi surface, $v_\alpha$ could be very large (small) for $e_\alpha \ll e_F$ ($e_\alpha \gg e_F$) orbits: typically $v_\alpha \approx 10$ ($v_\alpha \approx 0.01$) near $e_\alpha = e_F - 20$ MeV ($e_\alpha = e_F + 20$ MeV). Then the sign problem may arise when recursively computing $\chi_N^{[\alpha]}$ by Eq. (\ref{n_ave}).

Most numerical softwares run very quickly using numbers of double-precision floating-point format. (Because usually the machine-precision is double-precision.) So in practice, if we use double-precision numbers, how large the model space could be?
Our experience is that there is no sign problem up to the case of $2N = 24$ particles in a single-particle space of dimension $D = 2\Omega = 48$. (The model space is half-filled; the particle number is the largest considering the particle-hole symmetry \cite{Talmi_1982, Jia_2016}. $\Omega$ is the number of vacancies for Kramers pairs.) In this case Matlab {\emph{fminunc}} costs typically $0.3$ seconds to minimize $\bar{E}$, on a laptop by serial computing (not in parallel).

The modern mean-field theory uses large model spaces (for example, $15$ harmonic-oscillator major shells). In this case double-precision is not enough, and we resort to softwares that run quickly using arbitrary-precision numbers, for example, Mathematica. In principle, any algorithm running into the sign problem could use this strategy of increasing precision. However in practice, computing with arbitrary-precision numbers is usually much slower than with double-precision numbers; so the formulas for coding must be simple so that the total computer time cost is feasible.


The VDPC algorithm is designed to increase the valence space gradually: first minimizes $\bar{E}$ in a small valence space (of dimension $2\Omega \approx 50$ to use double-precision) around $e_F$ to quickly get the big picture, next refines the solution in larger valence spaces until the desired convergence. The Nilsson levels below the valence space are completely filled and form an inert core, the core simply corrects the valence-space single-particle energy by its HF mean field (\ref{e_HF}). Specifically, the algorithm has five steps:

1. We sort the single-particle basis states $|\alpha\rangle$ by their HF energy $e_\alpha$ (\ref{e_HF}), and occupy the lowest $2N$ basis states. In other words, we solve the HF equation but without mixing the basis states. (This work is VDPC+BCS, not VDPC+HFB.) This step is not needed if the input single-particle basis is already the HF basis. 2. We select around $e_F$ the first valence space (VS1) of dimension $2\Omega \approx 50$ (to use double-precision). Then we input both the energy [Eq. (\ref{H_ave_1})] and the gradient [Eqs. (\ref{grad_v_1}) and (\ref{grad_v_2})] into Matlab {\emph{fminunc}}, to quickly minimize $\bar{E}$. The resultant $v_\alpha$ of VS1 is called $v^{(1)}_\alpha$. 3. We select around $e_F$ the second valence space (VS2) of dimension $2\Omega \approx 200$. We initialize $v_\alpha$ of VS2 to be $v^{(1)}_\alpha$ if $|\alpha\rangle$ belongs to VS1, and to be a very large (small) number if $|\alpha\rangle$ is not in VS1 and $e_\alpha < e_F$ ($e_\alpha > e_F$) so that $n_\alpha \approx 1$ ($n_\alpha \approx 0$). Then we use the analytical formulas (\ref{v_1}) and (\ref{v_2}) to iterate $v_\alpha$ until convergence (usually 10 iterations are enough). The resultant $v_\alpha$ of VS2 is called $v^{(2)}_\alpha$. VS2 is large enough so that $v^{(2)}_\alpha$ is very close to the final solution. 4. For all the basis states $|\beta\rangle$ not in VS2, estimate $v_\beta$. We substitute $v^{(2)}_\alpha$ into the asymptotic expressions (\ref{v_L}) and (\ref{v_H}) to compute $v_\beta$. (This is the 1st order perturbation: determine $v_\beta$ from $v^{(2)}_\alpha$ of VS2.) Next we substitute $v^{(2)}_\alpha$ and $v_\beta$ into Eqs. (\ref{v_L}) and (\ref{v_H}) again, to compute the final $v_\beta$, labeled as $v^{{\rm{est}}}_\beta$. (This is the 2nd order perturbation: consider corrections from mutual interactions among $v_\beta$.) The corresponding occupation number is $n^{{\rm{est}}}_\beta$. 5. Choose two cutoffs $n_{\min}$ and $n_{\max}$ (for example $n_{\min} = 10^{-6}$ and $n_{\max} = 1 - 10^{-6}$), and select the third valence space (VS3): VS3 consists of VS2 and those basis states $|\beta\rangle$ satisfying $n_{\min} \le n^{{\rm{est}}}_\beta \le n_{\max}$. We initialize $v_\alpha$ of VS3 to be $v^{(2)}_\alpha$ if $|\alpha\rangle$ belongs to VS2, and to be $v^{{\rm{est}}}_\beta$ if $|\alpha\rangle$ is not in VS2. Then we use the analytical formulas (\ref{v_1}) and (\ref{v_2}) to iterate $v_\alpha$ in VS3 until the desired convergence. The resultant $v_\alpha$ of VS3 is called $v^{(3)}_\alpha$. This finishes the VDPC algorithm.

The next section demonstrates the algorithm in a semi-realistic example, giving the actual time cost and energy-convergence pattern.


\section{Realistic Example  \label{Sec_example}}

In this section we apply the VDPC + BCS algorithm to the semi-realistic example of the rare-earth nucleus $^{158}_{~64}$Gd$_{94}$. (This example has been used in our recent paper \cite{Jia_2017} on deformed generalized seniority.) The purpose is to demonstrate the effectiveness of the algorithm under realistic interactions. For simplicity, we consider only the neutron degree of freedom, governed by the anti-symmetrized two-body Hamiltonian
\begin{eqnarray}
H = \sum_{\alpha} \epsilon_\alpha a_\alpha^\dagger a_\alpha + \frac{1}{4} \sum_{\alpha \beta \gamma \delta} V_{\alpha \beta \gamma \delta} a_\alpha^\dagger a_\beta^\dagger a_\gamma a_\delta . \label{H_example}
\end{eqnarray}
The single-particle levels $\epsilon_\alpha$ are eigenstates of the Nilsson model \cite{Nilsson_1955}. The Nilsson parameters are the same as in Ref. \cite{Jia_2017}, here we only repeat $\beta = 0.349$ (the experimental quadrupole deformation \cite{NNDC}). The neutron residual interaction $V_{\alpha \beta \gamma \delta}$ is the low-momentum {\emph{NN}} interaction $V_{{\rm{low}}{\textrm{-}}k}$ \cite{Bogner_2003} derived from the free-space N$^3$LO potential \cite{Entem_2003}.

Specifically, we use the code distributed by Hjorth-Jensen \cite{Morten_code} to compute (without Coulomb, charge-symmetry breaking, or charge-independence breaking) the two-body matrix elements of $V_{{\rm{low}}{\textrm{-}}k}$ in the spherical harmonic oscillator basis up to (including) the ${\mathcal{N}} = 14$ major shell, with the standard momentum cutoff $2.1$ fm$^{-1}$. (${\mathcal{N}} = 2n_r+l$ is the major-shell quantum number.) Then the Nilsson model is diagonalized in this spherical ${\mathcal{N}} \le 14$ basis, the eigen energies are $\epsilon_\alpha$ and the eigen wavefunctions transform the spherical two-body matrix elements into those on the Nilsson basis as used in the Hamiltonian (\ref{H_example}).

This procedure has several assumptions. Mainly the proton-neutron interaction generates the static deformation and self-consistently the Nilsson mean-field. The residual proton-neutron interaction is neglected, and in the Hamiltonian (\ref{H_example}) the part of the neutron-neutron interaction already included in the Nilsson mean field $\epsilon_\alpha$ is not removed from $V_{\alpha \beta \gamma \delta}$. These assumptions make the example semi-realistic. Our goal is to demonstrate the effectiveness of the VDPC algorithm, not to accurately reproduce the experimental data.

\begin{figure}
\includegraphics[width = 0.5\textwidth]{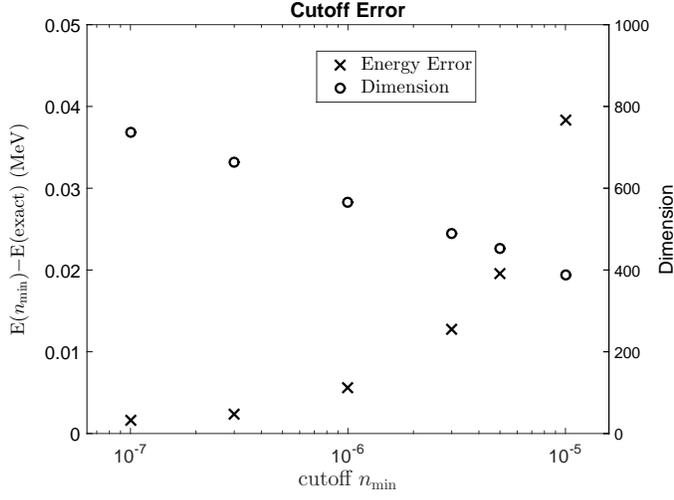}
\caption{\label{Fig_Ed_n} The cutoff error and dimension of six different model spaces (VS3). The horizontal axis shows the cutoff $n_{\min}$, so the model space consists of all Nilsson levels $\alpha$ satisfying $n_\alpha \ge n_{\min}$. The cross symbols correspond to the left axis and show the cutoff error: the energy of each model space $E(n_{\min})$ relative to the exact energy of the full space $E({\rm{exact}})$. The circle symbols correspond to the right axis and show the dimension of each model space. }
\end{figure}

\begin{figure}
\includegraphics[width = 0.5\textwidth]{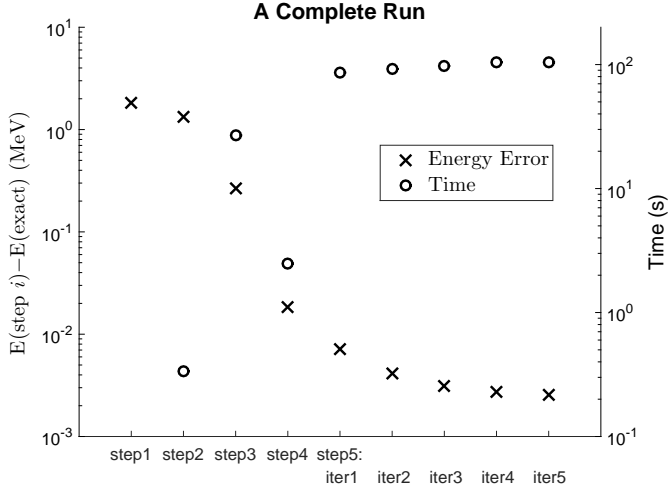}
\caption{\label{Fig_completeRun} Energy and time in a complete run. The horizontal axis shows the $5$ steps listed in Sec. \ref{Sec_code}, where step 5 is divided into $5$ iterations. The cross symbols correspond to the left axis and show the energy error after each step or iteration, relative to the final exact minimum $E({\rm{exact}})$ (converged energy in the full space). The circle symbols correspond to the right axis and show the computer time spent by each step or iteration. }
\end{figure}

\begin{figure}
\includegraphics[width = 0.5\textwidth]{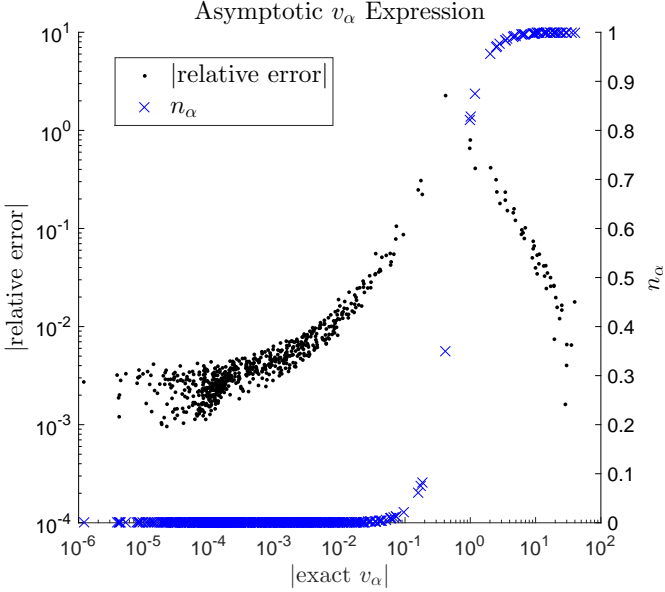}
\caption{\label{Fig_vn} (Color online) The error of the asymptotic $v_\alpha$ expression relative to the exact $v_\alpha$ expression. The relative error $RE \equiv (v_\alpha^{\rm{asymptotic}}/v_\alpha^{\rm{exact}}) - 1$. The horizontal axis shows $|v_\alpha^{\rm{exact}}|$, the absolute value of the exact $v_\alpha$. The dot symbols correspond to the left vertical axis and show $|RE|$. The cross symbols correspond to the right vertical axis
and show the exact occupation number $n_\alpha$. }
\end{figure}

All the numerical calculations of this work were done on a laptop. The laptop has one quad-core CPU (Intel Core i7-4710MQ @ 2.5 GHz), but we used only serial computing on a single core (no parallel computing). All time costs plotted in the figures or given in the text are the actual time costs spent on this laptop. There is only one exception: the full-space calculation (including all Nilsson levels after diagonalizing in the spherical ${\mathcal{N}} \le 14$ basis) was done on a server computer (also by serial computing) because of memory, which provides $E({\rm{exact}})$ in Figs. \ref{Fig_Ed_n} and \ref{Fig_completeRun} and data in Fig. \ref{Fig_vn}. This work uses Matlab R2015a and Mathematica 10.2, to give the actual software version.

\begin{figure}
\includegraphics[width = 0.5\textwidth]{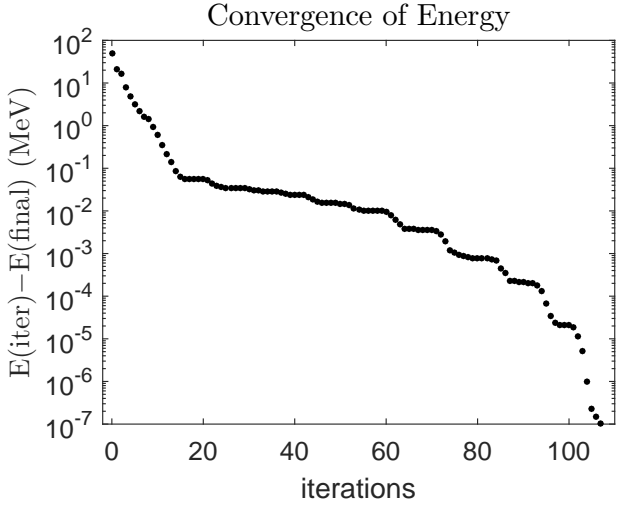}
\caption{\label{Fig_E_matlab} Convergence of energy in VS1 by {\emph{Matlab}}. The horizontal axis shows the number of iterations. The vertical axis shows the energy at each iteration E(iter), relative to the final converged energy E(final). }
\end{figure}

We follow the steps listed in Sec. \ref{Sec_code}. In step 1 we sort the Nilsson basis by their HF energy $e_\alpha$ (\ref{e_HF}). In step 2 we select VS1, and use {\emph{Matlab}} {\emph{fminunc}} to minimize $\bar{E}$. VS1 consists of all Nilsson levels $\alpha$ satisfying $e_F - 5.38 {\rm{MeV}} < e_\alpha < e_F + 5.5 {\rm{MeV}}$. It has dimension $2\Omega = 48$ ($24$ Nilsson levels below $e_F$ and $24$ above $e_F$). Starting from a random initial $v_\alpha$, {\emph{Matlab}} quickly minimizes $\bar{E}$ in about $0.3$ second. This process is plotted in Fig. \ref{Fig_E_matlab}: how the energy converges as the number of iterations increases.

\begin{figure}
\includegraphics[width = 0.5\textwidth]{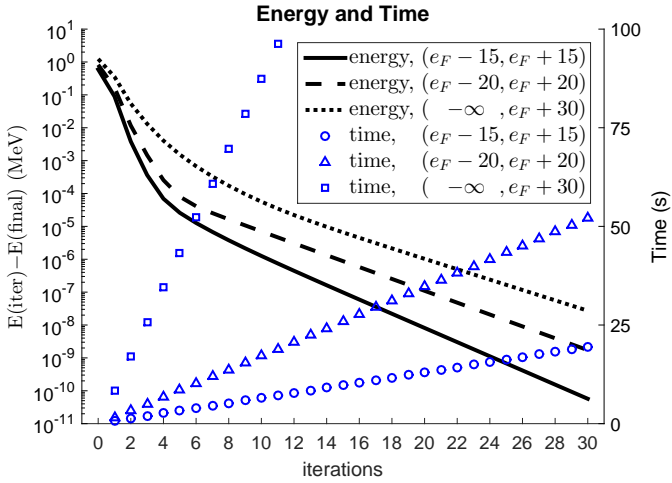}
\caption{\label{Fig_E_Math1} (Color online) Energy and time in three different model spaces (VS2) by {\emph{Mathematica}}. The model space $(e_F-15,e_F+15)$ consists of all Nilsson levels $\alpha$ satisfying $e_F - 15 {\rm{MeV}} < e_\alpha < e_F + 15 {\rm{MeV}}$. ``$-\infty$'' means including all Nilsson levels below the Fermi energy $e_F$. The solid, dashed, dotted lines correspond to the left vertical axis and show the energy errors in the three model spaces. The circle, triangle, square symbols correspond to the right vertical axis and show the accumulated computer time cost after each iteration. All time costs in this work refer to that by serial computing on a laptop (CPU is Intel Core i7-4710MQ @ 2.5 GHz, no parallel computing).}
\end{figure}

In step 3 we select VS2, and use {\emph{Mathematica}} to minimize $\bar{E}$ by iterating Eqs. (\ref{v_1}) and (\ref{v_2}). Figure \ref{Fig_E_Math1} plots the energy and time in three different choices for VS2. The accumulated computer time cost increases linearly with the number of iterations, so each iteration costs the same time approximately. The energy error decreases the fastest in the first few iterations (by several orders of magnitude). Overall, the curve is linear on the log-scale plot, so energy converges exponentially with the number of iterations. In this work we choose VS2 to be $(-\infty,e_F+20)$ and go to step 4, where we estimate $v_\alpha$ in the full space by the asymptotic expressions (\ref{v_L}) and (\ref{v_H}). Step 4 costs about $2.5$ seconds.

\begin{figure}
\includegraphics[width = 0.5\textwidth]{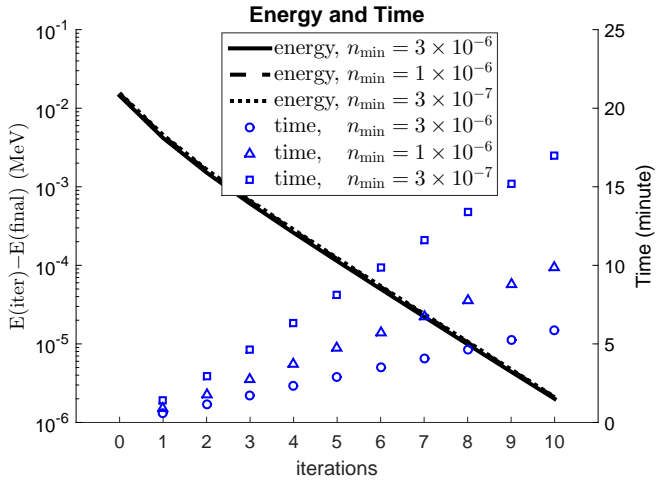}
\caption{\label{Fig_E_Math3} (Color online) Energy and time in three different model spaces (VS3) by {\emph{Mathematica}}. The model space $n_{\min}$ consists of all Nilsson levels $\alpha$ satisfying $n_\alpha \ge n_{\min}$. Other symbols have similar meanings to those of Fig. \ref{Fig_E_Math1}. The solid, dashed, dotted lines overlap and are indistinguishable. }
\end{figure}

In step 5, we select VS3 by choosing two cutoffs $n_{\min}$ and $n_{\max}$. There are $94$ Nilsson levels below $e_F$, and in this work we include all of them by setting $n_{\max} = 1$. Six choices of $n_{\min}$ generate six different VS3, their dimension and cutoff error (relative to the full space when $n_{\min} = 0$) are shown in Fig. \ref{Fig_Ed_n}. Within each VS3, the computer time cost and convergence of energy are plotted in Fig. \ref{Fig_E_Math3}. After only $2$ iterations the energy has converged within $2$ keV, so practically we need very few iterations in step 5.

In summary, Fig. \ref{Fig_completeRun} combines the above five steps and shows a complete run. Step 1 is HF and costs negligible time. The pairing correlation energy is $1.83$ MeV (defined as the energy difference between the HF Slater determinant and the final coherent pair condensate). Step 2 reduces the energy error to $1.33$ MeV in $0.335$ second. Step 3 uses $10$ iterations and reduces to $267$ keV in $27$ seconds. Step 4 reduces to $18$ keV in $2.5$ seconds. Step 5 uses the cutoff $n_{\min} = 3 \times 10^{-7}$ and costs the largest time. The error gradually reduces to $2.4$ keV, which is the cutoff error corresponding to $n_{\min} = 3 \times 10^{-7}$. Figure \ref{Fig_completeRun} is just an example; for a desired accuracy, fine-tuning parameters in these five steps finds the shortest time cost. In passing, extending to VDPC+HFB, we may not need the slowest step 5 when computing $v_\alpha$ on each intermediate canonical basis, because after step 4 the energy error is already pretty small ($18$ keV). Only at the final iterations step 5 was needed to reach complete convergence.

The asymptotic expressions (\ref{v_L}) and (\ref{v_H}) very well reproduce the exact $v_\alpha$ (\ref{v_1}) and (\ref{v_2}) away from the Fermi surface. Figure \ref{Fig_vn} compares them at the energy minimum in the full space. The horizontal axis shows $|v_\alpha^{\rm{exact}}|$ instead of $v_\alpha^{\rm{exact}}$, because some $v_\alpha^{\rm{exact}}$ are negative with very small magnitudes. (The range is $-2.40 \times 10^{-4} \le v_\alpha^{\rm{exact}} \le 39.7$). Near the Fermi surface, the asymptotic expressions (\ref{v_L}) and (\ref{v_H}) are not justified so $|RE|$ is big ($|RE|$ is the absolute value of relative error). Going away from the Fermi surface, $|RE|$ becomes smaller and smaller. There are $680$ different $v_\alpha$ (the full-space dimension is $1360 = 680 \times 2$), $661$ of them have $|RE| < 10\%$, $568$ of them have $|RE| < 1\%$. Figure \ref{Fig_vn} shows $v_\alpha$ at the energy minimum; near the minimum $|RE|$ has a similar pattern, which makes step 4 effective.

\begin{figure}
\includegraphics[width = 0.5\textwidth]{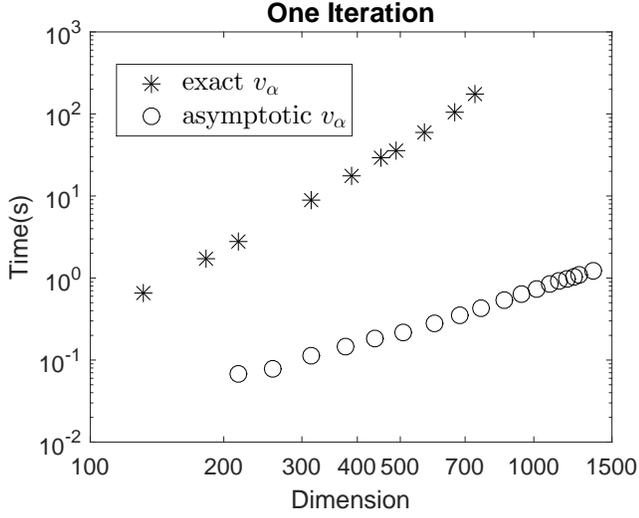}
\caption{\label{Fig_t_D} Computer time cost of one iteration in different model spaces. The horizontal axis shows the dimension of each model space, and the vertical axis shows the time cost (averaged over many iterations). The asterisk symbols iterate by the exact $v_\alpha$ expressions (\ref{v_1}) and (\ref{v_2}), and the circle symbols iterate by the asymptotic $v_\alpha$ expressions (\ref{v_L}) and (\ref{v_H}). }
\end{figure}

The new algorithm minimizes $\bar{E}$ through iterating $v_\alpha$, by the exact expressions (\ref{v_1}) and (\ref{v_2}) or the asymptotic expressions (\ref{v_L}) and (\ref{v_H}). The computer time cost per each iteration mainly depends on the dimension of the (single-particle) model space. Figure \ref{Fig_t_D} shows that this time increases approximately linearly with dimension on the log-log plot. We perform a linear least-squares fit in the form $\log(T) = \alpha \log(D) + C$ ($T$ is time in unit of second, $D$ is dimension, $\alpha$ and $C$ are fitting parameters). The result is $T = (D/155)^{3.19}$ for the exact $v_\alpha$, and $T = (D/1234)^{1.62}$ for the asymptotic $v_\alpha$. The latter is much quicker.

\begin{figure}
\includegraphics[width = 0.5\textwidth]{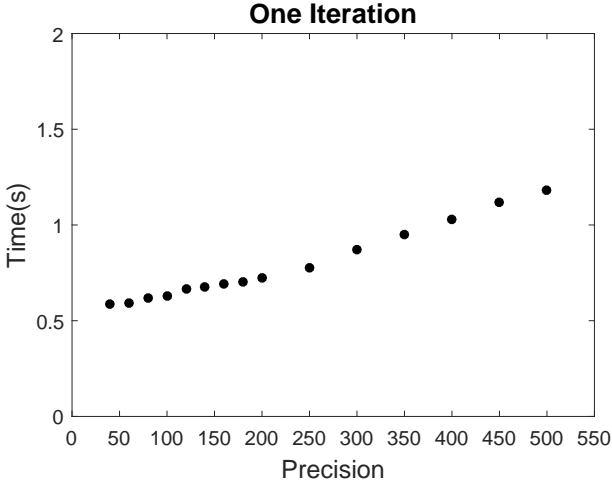}
\caption{\label{Fig_time_prec} Computer time cost of one iteration using different precision. The horizontal axis shows the precision, and the vertical axis shows the time cost [averaged over many iterations, by the exact $v_\alpha$ expressions (\ref{v_1}) and (\ref{v_2})]. The model space has dimension $132$. }
\end{figure}

The new algorithm needs arbitrary-precision numbers in large model spaces, when the sign problem arises using double-precision numbers. Usually computing with arbitrary-precision is slower than that with double-precision; however the new formulas (\ref{n_ave}), (\ref{v_1}), (\ref{v_2}) are simple so that the total time cost is feasible. We use $120$ effective digits (by {\emph{Mathematica}} function SetPrecision[120]) for all the calculations in this work, which guarantees no sign problem. Using fewer effective digits to reduce time cost is possible as shown in Fig. \ref{Fig_time_prec}. In a model space of dimension $132$ [$(e_F-15,e_F+15)$ of Fig. \ref{Fig_E_Math1}], we use from $40$ to $500$ effective digits --- all of them guarantee no sign problem. The time cost increases, but the slope is rather small. The small slope is a feather of {\emph{Mathematica}}, and for this reason we do not fine-tune precision in this work but always use $120$ effective digits. If using another software with a big slope, fine-tuning precision should be worthwhile.

Finally we suggest some directions to further optimize the algorithm. First, Fig. \ref{Fig_completeRun} shows that step 5 costs the most time, in fact computing $\langle \phi_N^{[\alpha]} | H | \phi_N^{[\alpha]} \rangle$ and $\langle \phi_{N-1}^{[\alpha]} | H | \phi_{N-1}^{[\alpha]} \rangle$ in Eqs. (\ref{v_1}) and (\ref{v_2}) is very time-consuming. Currently we use only serial computing on a single core; an easy speedup would be computing $\langle \phi_N^{[\alpha]} | H | \phi_N^{[\alpha]} \rangle$ and $\langle \phi_{N-1}^{[\alpha]} | H | \phi_{N-1}^{[\alpha]} \rangle$ in parallel, by distributing each of them (each $\alpha$) to different cores. Second, in large model spaces (for example ${\mathcal{N}} \le 14$ of this work) majority of $\alpha$ orbits are above the Fermi surface and computed by Eq. (\ref{v_2}). For those $e_\alpha \gg e_F$ orbits, a very good approximation [better than Eq. (\ref{v_H})] to Eq. (\ref{v_2}) would be replacing $\langle \phi_{N-1}^{[\alpha]} | H | \phi_{N-1}^{[\alpha]} \rangle$ by $\langle \phi_{N-1} | H | \phi_{N-1} \rangle$. If this approximation caused little error in the final average energy, it should be used to greatly reduce the time cost. Third, Fig. \ref{Fig_completeRun} shows that step 4 [iterates asymptotic $v_\alpha$ expressions (\ref{v_L}) and (\ref{v_H})] is very quick and very effective, it could be used many times at different places.

Our results suggest that the realistic $V_{{\rm{low}}{\textrm{-}}k}$ interaction does not cause divergences in the pairing channel. In this work the full space (${\mathcal{N}} \le 14$) has dimension $1360$. Figure \ref{Fig_Ed_n} shows that in the truncated subspace of dimension 452 ($n_{\min} = 5 \times 10^{-6}$), the energy cutoff error is already less than $20$ keV --- energy has converged, roughly speaking. This also suggests truncating the space by occupation numbers may be more effective than that by single-particle energies. In some cases, a few tens of keV may be important \cite{Lunney_2003}, then the tiny occupation numbers (for example smaller than $5 \times 10^{-6}$) can not be neglected and we should further enlarge the subspace.

\section{Conclusions  \label{Sec_summary}}

This work proposes a new algorithm to perform the variational principle directly on the coherent pair condensate (VDPC). It always conserves the particle number, and avoids the time-consuming particle-number projection by gauge-angle integration. Specifically, we derive analytical expressions for the average energy and its gradient in terms of the coherent pair structure $v_\alpha$. Requiring the gradient vanishes, we obtain the analytical expression of $v_\alpha$ at the energy minimum. The new VDPC algorithm iterates this $v_\alpha$ expression to minimize energy until practically arbitrary precision. We also find the asymptotic expression of $v_\alpha$ that is highly accurate (see Fig. \ref{Fig_vn}) and numerically very fast (see Fig. \ref{Fig_t_D}). These analytical expressions look quite simple and allow easy physical interpretations.

We demonstrate the new VDPC algorithm in a semi-realistic example using the realistic $V_{{\rm{low}}{\textrm{-}}k}$ interaction and large model spaces (up to $15$ harmonic-oscillator major shells). The energy-convergence pattern and actual computer time cost are given in detail. Figure \ref{Fig_completeRun} shows an example run from beginning to end. How to organize the analytical results of this work into an optimal numerical algorithm remains an open question, and some suggestions are given at the end of Sec. \ref{Sec_example}.

It is a good property of a specific interaction to converge naturally in the pairing channel; otherwise a phenomenological cutoff is needed to truncate the pairing-active model space. The zero-range pairing, frequently used together with the Skyrme density functional, does not have this property. The Gogny force has this good property. Our results show that the realistic $V_{{\rm{low}}{\textrm{-}}k}$ interaction has this good property. However, tiny occupation numbers contribute to the energy (see Fig. \ref{Fig_Ed_n}), thus should be kept if the desired accuracy is high.

This work considers VDPC+BCS. Extending to VDPC+HFB needs the gradient of the average energy with respect to changes of the canonical single-particle basis. In Sec. \ref{Sec_grad_basis} we have derived them analytically and transformed the minimization into a diagonalization problem; VDPC+HFB will be the topic of our next work in the series.

\section{Acknowledgements}

Support is acknowledged from the National Natural Science
Foundation of China No. 11405109.

%
%
%
%
%
%
%

\newpage

\end{document}